\documentclass[12pt]{iopart}
\usepackage{iopams}
\usepackage{graphicx,cite,color,epsf}

\usepackage{graphicx,epsf}
\usepackage{hyperref}
\usepackage{xcolor}
\usepackage{psfrag}
\begin{document}

\title{Coagulation-flocculation process on a lattice: Monte Carlo simulations}

\author{Viktoria Blavatska$^{1,2}$\footnote{Author to whom any correspondence should be addressed}, Jaroslav Ilnytskyi$^{1,3}$, and Erkki L\"ahderanta$^{4,5}$ }
\address{$^1$
Institute for Condensed
Matter Physics of the National Academy of Sciences of Ukraine,
79011 Lviv, Ukraine}
\address{$^2$
Dioscuri Centre for Physics and Chemistry of Bacteria,
Institute of Physical Chemistry, Polish Academy of Sciences, 01-224 Warsaw, Poland}
\address{$^3$
Institute of Applied Mathematics and Fundamental Sciences, 
Lviv Polytechnic National University, 12 S. Bandera Str., UA-79013 Lviv, Ukraine}
\address{$^4$ 
Department of Physics, School of Engineering Science,
LUT University, Yliopistonkatu 34, FI-53850 Lappeenranta, Finland}
\address{$^5$
Department of Physics, Universitat de les Illes Balears, Cra Valldemossa, km. 7.5, 07122, Palma, Spain 
}

\ead{viktoria@icmp.lviv.ua}

\begin{abstract}

Coagulation-flocculation, the physicochemical process widely used for purification a wastewater, is affected both by chemical details of involved polymers and by the statistics of their conformations on a large scale. The latter aspect is covered in this study by employing a coarse-grained modelling approach based on a combination of two paradigms of statistical mechanics. One is the self-avoiding walk (SAW) which generates a range of conformations for a linear polymer of $N_{{\rm SAW}}$ monomers. Another one is a non-trivial diffusion limited aggregation (DLA) process of $N_{{\rm  DLA}}$ impurities (referred thereafter as ``particles") which describes their coagulation occurring with the probability $0< p \leq 1$ ($p=1$ recovers a standard DLA). DLA of diffusive particles is complemented by their irreversible adsorption on the SAW monomers occurring with the probability equal to one, both processes resulting in formation of the DLA-SAW agglomerates. The dynamics of formation of such agglomerates, as well as their fractal dimensions and internal structure are of practical interest. We consider a range of related characteristics, such as: (i) absolute $N_a$ and relative $n_a$ adsorbing efficiencies of SAW; (ii) effective gyration radius $R_{g {\rm DLA-SAW}}$ of the DLA-SAW agglomerates; and (iii) the fractal dimension $D_{{\rm DLA-SAW}}$ of these aggregates. These are studied within a wide range for each parameter from a set $\{p,N_{{\rm  DLA}},N_{{\rm SAW}}\}$.
\end{abstract}
\submitto{Journal of Physics A: Mathematical and Theoretical}
\date{\today}
\maketitle

\section{Introduction}\label{sec:1}

Coagulation-flocculation is an important physicochemical process to purify a wastewater off impurities \cite{Sahu2013,Daud2023}. A coagulant, typically a short polymer termed often as a ``clarifying agent", neutralizes the particles’ charge allowing them to coagulate. A flocculant, added to wastewater, aids further conglomeration of particles into larger agglomerates, speeding up the process of sedimentation \cite{Lee2014,Eltaweel2023}. High molecular weight flocculants exhibit bridging flocculation, when a single molecule adsorbs a number of particles resulting in formation of a necklace-like structure \cite{Hogg2013}. Polymers commonly used in applications with inorganic solids such as clays and silts are anionic, whereas cationic polymers are used to settle organic solids such as animal waste or vegetation. They can be either synthetic, e.g. alum, lime, ferric chloride, polyaluminium, or derived from plant parts \cite{Bandala2013, Daud2018, Badrus2018}, bacteria \cite{AbuHasan2021, Yee2021}, and chitosan \cite{Iber2021}.

{In most wastewater purification protocols, coagulation-flocculation is an in\-ter\-me\-diate step, prior to the membrane filtration \cite{Zhao2021}. In this case both processes take place in bulk. However, there are many situations when the target particles for aggregation reside on an interface, e.g. the liquid-liquid, water-air, etc. \cite{Williams1992, Naganawa2012, Turpin2023, Lamanna2023}. Their aggregation is desirable either for the consequent removal of aggregates, or for the sake of the interface stabilization and control of its properties \cite{Guzman2021}. In this case the problem, obviously, reduces to the case of the two-dimensional one. Therefore, both the $3D$ and $2D$ cases are equally important from the application point of view, but in this study we consider the $2D$ case only.} 

The high molecular weight polymers used in bridging flocculation process are typically  of linear architecture, commonly based on polyacrylamide. In particular, Gibson et al. examined seven of such polyacrylamide polymers and they found that the two of them, Drewfloc 2449 and 2468, demonstrate the highest solid removal efficiency \cite{Gibson2020}. In general, adsorption mechanism for such flocculation is considered to be based on hydrogen bonding between the amide or hydroxyl groups of a polymer and hydroxylated sites on particles' surfaces \cite{Hogg2013}. 

Modelling the coagulation-flocculation faces serious difficulties as it is intrinsically a multiscale process involving events ranging from an atomistic to a macroscopic scale.  Indeed, interactions between particles and monomers of a polymer chain, be it a coagulant or flocculant, depends strongly on the chemical details of both. This aspect can be tackled by \textit{ab initio} and/or atomistic molecular dynamics simulations. The aim of the current study is to cover an opposite, macroscopic, side of the length scales spectrum. Firstly, the coagulant is considered in an implicit way, where we just assume that due to its presence the particles are able to coagulate with the tunable probability $0<p<1$. A linear high molecular weight flocculant is considered explicitly, assuming that it is able to adsorb particles irreversibly, with the probability equal to one. Our main focus will be on the macroscopic properties such as: (i) the efficiency of a flocculant to agglomerate particles; (ii) effective dimensions of agglomerates; and (iii) their fractal dimension. All these characteristics define hydrodynamic behavior of agglomerates during their removal from the wastewater via membrane filtering, gel chromatography, or other approaches \cite{Rashid2021,Shekho2024}. {We should mention that the modelling of the flocculation as a fractal DLA process have been done before \cite{Liu2017}. Both $2D$ and $3D$ DLA were considered but for the case of a point seed only.}

The description of the mixed solution, containing a polymeric flocculant and impurity particles, can be performed on required large scale level by employing the lattice models of polymers \cite{deGennes1979, desCloizeaux1982}. { In these terms, a flocculant is represented via the SAW of the size $N_{{\rm SAW}}$ \cite{Freed1981, Hooper2020}. On the other hand, particles aggregation can be described by the DLA process of the size $N_{{\rm DLA}}$ \cite{Platt1995}.} The concept of the DLA was initially introduced  \cite{Witten1981} as a discrete model of growth of dendritic clusters formed by irreversible aggregation of small particles. Since then, it became a paradigm in description of growth phenomena and pattern formation \cite{Halsey2000,Tomchuk2023}, in particular as a model of colloidal particle aggregation \cite{Antonsson2024,Lazzari2016}, growth of neuronal trees \cite{Luczak2010}, retinal arterial and venous vessels \cite{Lorthois2010},  thin film nucleation growth process\cite{Zhang2018},  growth of microbial colonies \cite{Tronnolone2018} etc.

Experimental studies reveal proportionality between the particles sedimentation rate and the flocculant molecular weight \cite{Li2009,Lu2016}. An explanation is given in terms of stronger adhesion, e.g., via hydrogen bonding, between the particles and the abundant hydroxyl groups in the case of the high molecular weight glycopolymers \cite{Lu2016}. The result, however, might depend on the chemical composition of a flocculant. It is of interest to clarify the effect of the flocculant molecular weight, $N_{{\rm SAW}}$, originated purely from the conformational statistics of long polymers. Other parameters affecting the flocculation rate that can be pointed out are: the coagulation probability $p$, and concentration of the particles, provided by $N_{{\rm DLA}}$.

We should note that, although both the SAW and DLA paradigms are well established and understood by now, combining them towards description of particular physical phenomena has not been exploited to its full extent yet. By doing this, one goes one step further from more simple cases of point-like seeds or the seed cores with a regular shape. Indeed, the seed core in a form of a SAW presents a complex structure that can adopt a wide range of conformations, from stretched to highly coiled ones, affecting the process of the DLA cluster growth in each case. By altering the probability of coagulation, $p$, one may shift an emphasis from aggregation of particles to their diffusion and \textit{vice versa}, steering the process towards some specific spatial structure of a cluster.

{Therefore, here we link two paradigms of DLA and SAW to model coagulation-flocculation on a large scale level. In doing this, DLA describes coagulation of particles, whereas SAW represents a linear high molecular weight flocculant in a good solvent. We may note that the same model is also applicable to other related problems, such as: adsorption of colloidal particles from a dilute aqueous suspension \cite{Shull1995, Prakash2015}; nanoparticle diffusion and adsorption in polymer melts \cite{Mller1997, Griffin2016, Dementeva2005, Zare2016}, etc. Finally, we would like to address the issue of relative scales of a flocculant and the impurity particles in the lattice type of modelling. The dimensions of flocculant monomers and of a particle are both defined via a lattice constant $a$. Since a polymer is modelled as a SAW, $a$ is lower bound by the persistent length $l_p$ of a polymeric flocculant, with the typical values of $l_p=0.5-1\textrm{nm}$. The scale $a$, however, is not upper bound because of the self-similarity feature of a polymer chain at $a>l_p$. Therefore, the monomer can be interpreted as a subchain with essentially larger dimensions than $l_p$, bringing the intrinsic length scale of a problem into the realms of tens of nanoseconds or further up. More close comparison of the scales with particular real life polymers is problematic and, in general, is not intended.} 

The layout of the paper is as follows. In the next Section \ref{II}, we give details of the computational methods used to construct the SAW and DLA clusters, 
the results are presented and discussed in Section \ref{III}, followed by Conclusions.

\section{Models and algorithms}\label{II}

The growth of the SAW is performed sequentially: the $n$th monomer is generated at the random, but yet unoccupied, lattice site adjacent to the $(n-1)$th monomer. The process continues until the required number of monomers, $N_{{\rm SAW}}$, is created. The model is known to capture perfectly the universal configurational properties of long, flexible polymer chains in a good solvent \cite{desCloiseaux} and has been studied in detail by both computer simulations \cite{Rosenbluth1955, 
MacDonald2000,Blavatska2008} and analytical approaches \cite{
Guillou1985, Guida1998}.  

The DLA process is typically initiated from a single ``seed'' particle. The second particle diffuses from a large distance away from a seed until it reaches it, in which case two particles merge into a cluster serving as a new seed for subsequent iterations. The process continues until the cluster of desired size $N_{{\rm  DLA}}$ is built. Since the perimeter sites of a cluster can be accessed more easily than those in its inner core, DLA cluster is characterized by a highly branched fractal structure. {Its fractal dimension $D_{{\rm  DLA}}$ is found from the relation connecting the cluster size $N_{{\rm  DLA}}$ and its effective linear dimensions given by the gyration radius ${R_g}_{{\rm  DLA}}$}
\begin{equation}
{\overline {R_g} }_{{\rm  DLA}} \sim N_{{\rm  DLA}} ^{1/D_{{\rm  DLA}}}, \label{dimension}
\end{equation}  
where the averaging ${\overline{ (\cdots)}}$ is performed over an ensemble of different DLA realizations. In general, the fractal dimension gives an estimate of the space-filling properties of  considered structure: the closer it is to the Euclidean dimension $D$ of embedding  space, the more ``dense'' is the structure. 

\begin{figure}[b!]
	\begin{centering}
\hspace{2.5cm}\includegraphics[width=65mm, height=50mm]{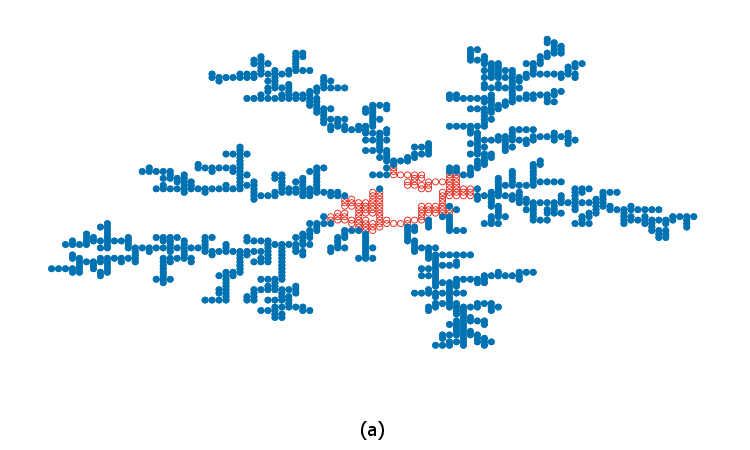}
\includegraphics[width=65mm, height=50mm]{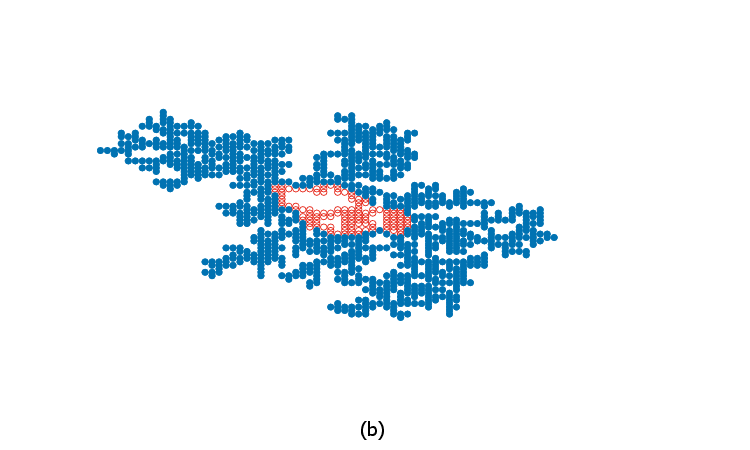}
  \caption {\label{cluster}(color online) The snapshots of SAW cluster of $N_{{\rm SAW}}=120$ particles (shown with red symbols) with $N_{{\rm  DLA}}=1000$ particles adsorbed during DLA process (shown with blue symbols) with sticking probabilities $p=1.0$ (a) and $p=0.1$ (b).}
\end{centering}
\end{figure}

{Note that the fractal dimension $D_{{\rm  DLA}}$, as defined by Eq. (\ref{dimension}), can be referred to as the {\it dynamical} one, describing growth of cluster dimensions $R_g$ with the increase of its size $N_{{\rm  DLA}}$. In  $D=2$, the value $D_{{\rm  DLA}}= 1.712(2)$ has been found \cite{Somfai2003}. Alternatively, $D_{{\rm  DLA}}$ can be obtained from the density correlation function for the completely grown DLA cluster, and smaller value, $D_{{\rm  DLA}}\approx 1.66$, is reported in this case \cite{Witten1981,Witten83a}.
}
Apart from a ``classic'' DLA,  the reaction limited cluster aggregation (RLCA) is also of interest \cite{Meakin1983,Jullien1984,Moran2021,Jungblut2019}. In this case, the 
non-vanishing repulsive forces between particles 
are taken into account, resulting in reduction of the aggregation rate, so that the aggregation (in our case, coagulation) probability is set lower than one. These two regimes, DLA and RLCA, can be attributed to the rapid and slow colloid aggregation, respectively, as defined in colloid science \cite{Verwey1948}. 
{ DLA with the non-trivial coagulation probability $0<p\leq 1$ of diffusive particles in $D=2$ has been analyzed in Ref. \cite{Ranguelov2012}. In the case of rapid aggregation, $p=1$, a highly porous ``classic'' DLA cluster is reproduced, see Fig. \ref{cluster} (a). In a moderately slow aggregation regime, $p<1$, the process became more diffusion driven allowing particles more time to explore the voids inside a fractal structure. As the result, more compact and dense DLA clusters are created that are characterized by the increase of $D_{{\rm  DLA}}$ with the decrease of $p$, see Fig. \ref{cluster} (b). In the extremely slow aggregation regime, at $p \to 0$, the limit of the Eden model is recovered with $D_{{\rm  RLCA}} = D$ \cite{Jullien1984}.} Recently, such modified DLA process has been used to simulate aggregation of wax and asphaltene particles in a crude oil \cite{Sun2017}, as well as in the antimicrobial peptide attack on supported lipid membranes \cite{Heath2018}.

When DLA is performed on SAW, the seed is not local and some generalizations of the standard DLA should be discussed.  Note also that SAW itself a fractal object, the exact value for its fractal dimension in $D=2$ is found to be $D_{{\rm SAW}}=4/3$ \cite{Nienhuis1982}. The multi-seed generalization of DLA has been considered in Ref.\cite{Sidoravicius2019}. At large separations between individual seeds, the crossover from a fractal to a uniform aggregate structure was observed at a certain length scale dependent on the concentration of mobile particles \cite{Witten1983, Sidoravicius2019}. Another generalization of DLA is obtained for the seed with a non-zero dimensions. As shown by Wu et al. \cite{Wu2013}, the fractal dimension of DLA cluster decreases with an increase of the dimensions of a seed particle. The same conclusion is derived from the studies of DLA growth on spherical surfaces of various radius \cite{Choi2010,Tenti2021}. Besides a seed of a spherical shape, its linear counterpart has also been considered. In particular,
the DLA on a linear seed, reproducing a fibre, was studied in Refs. \cite{Qiang2006, Procaccia2021}, where an increase of its length resulted in gradual decrease of $D_{{\rm  DLA}}$ towards $D=1$, the value of Eucledean dimension of a seed, and the range of singularities becomes narrower. More complex structure of a seeding core has been also examined \cite{14}. In this respect, { DLA in a SAW as a seed can be interpreted } as yet another extension of the standard DLA process.




{

To construct SAW  on  $D=2$ square lattice, we use the  pruned-enriched Rosenbluth Method (PERM) \cite{Grassberger97}, based on the  Rosenbluth-Rosenbluth (RR)  algoritm of growing chain \cite{Rosenbluth1955} and reinforced by enrichment strategies \cite{Wall59}. The first monomer is introduced at a random site of a lattice with the coordinates $x_1,y_1$. Each following $n$th monomer is added at a randomly chosen site adjacent to the $(n-1)$th one, such that its coordinates $\{x_n,y_n\}$ satisfy the conditions: $\{x_n=x_{n-1}\pm 1, y_n=y_{n-1} \} $ or $\{x_n=x_{n-1},y_n=y_{n-1}\pm 1\}$  ($n\leq N_{{\rm SAW}}$, where $N_{{\rm SAW}}$ is the total length of polymer  chain). 
The weights 
\begin{equation} W_n\sim \left(\prod_{l{=}2}^n m_l\right)^{-1} \label{weight1}\end{equation} 
are prescribed to each SAW configuration with $n$ monomers, where $m_l$ is the number of free lattice sites, where $l$th monomer could be potentially added. When the polymer chain of total length $N_{{\rm SAW}}$  is constructed, the new one starts from the same starting point, until the desired number of growing chain configurations are obtained. 

Population control in PERM suggests {\it pruning} configurations with too small weights, and {\it enriching} the sample with copies of the high-weight configurations \cite{Grassberger97}.  To this end, two thresholds values $W_n^{<}$ and $W_n^{>}$ are chosen depending on the running value of partition sum $Z_n=\sum_{{\rm conf}} W_n^{{\rm conf}}$, where summation is performed over existing configurations of a chain. If the current weight $W_n$ of an $n$-monomer chain is less than $W_n^{<}$,  the chain is either discarded with probability $1/2$, or it is kept and its weight is doubled.  If $W_n$ exceeds  $W_n^{>}$, the configuration is doubled and the weight of each copy is taken as half the original weight. The pruning-enrichment control parameters are adjusted in such a way that on average 10 chains of total length $N_{{\rm SAW}}$ are generated per each iteration \cite{Bachmann2004}. 

At the second stage of our simulation, the ensemble of constructed SAW clusters are used as seeds for DLA process with $N_{{\rm  DLA}}$ particles. We draw an imaginary circle of  radius $R=R_0+R_{max}$ \cite{Lee1992,Braga2011} (so-called birth circle, see Fig. \ref{R0}) around each constructed SAW cluster, where $R_{max}$ is  defined as the distance { between the center of DLA-SAW aggregate and the farthest adsorbed particle with respect to it}, and $R_0$ is chosen to be sufficiently large (we used the value of $R_0=50$). The model has been tested for different $R_0$ \cite{Lee1992}, and the increase of the value of this parameter leads to an increase of average time for diffusive particles to reach the perimeter of aggregate, but does not influence the quantitative characteristics of adsorption processes. 

Let us define the position of $i$th diffusive particle ($i=1,\ldots,N_{{\rm  DLA}}$) at time $t$ with its coordinates $\{X_i(t),Y_i(t)\}$.  Each particle starts to move from the point randomly chosen on the circle, so that the coordinates of its initial position satisfy the condition $X_i^2(0)+Y_i^2(0)=R^2$.
The particle is not allowed to cross the perimeter of a circle and moves only inside of it, so that at any time step $t$ we have: $X_i^2(t)+Y_i^2(t)\leq R^2$.
When a particle reaches the position such that any of its neighboring site contains a monomer of SAW, i.e. $\forall n$:
$\{X_i(t)\pm1=x_n, Y_i(t)=y_n\}$ or $\{X_i(t)=x_n, Y_i\pm1=Y_n \}$,
it is adsorbed with probability equal to one and a new particle starts to move from a new randomly chosen point on a birth circle. The following particles can be adsorbed not only to a SAW cluster, but also aggregate with the previously adsorbed particles, with a chosen probability $p$. The process is stopped when the desired number of particles $N_{{\rm  DLA}}$ are adsorbed.

\begin{figure}[t!]
	\begin{center}
\hspace{1.5cm}\includegraphics[width=65mm, height=80mm]{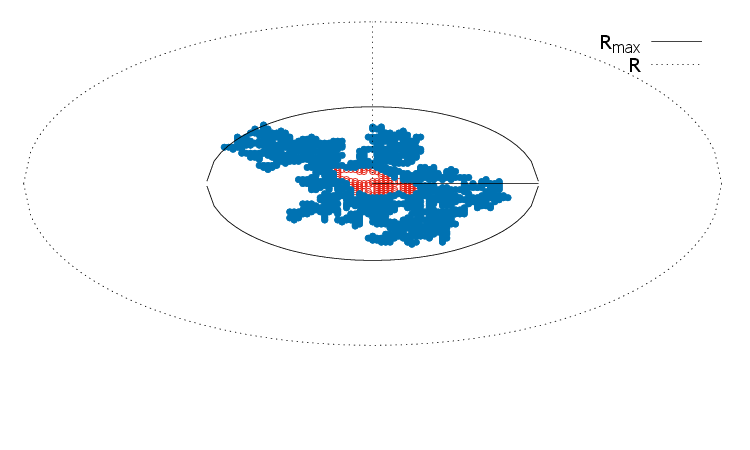}
  \vspace{-1.5cm}
  \caption {(color online) The snapshots of constructed DLA-SAW cluster with $N_{{\rm SAW}}=120$  (shown with red symbols) and $N_{{\rm  DLA}}=1000$ (shown with blue symbols) with sticking probability $p=0.1$,  surrounded by birth circle of the radius $R=R_0+R_{max}$. } 
\label{R0}	
\end{center}
\end{figure}

Note that the double averaging is needed to be performed for any observable of interest $O$ in the considered problem. 
First, we perform the averaging on a fixed SAW cluster over an ensemble of $M$ different DLA realizations 
\begin{equation}
  {\overline O } = \frac{1}{M}\sum_{i=1}^MO^i. \label{aver1}
 \end{equation}
  
Also the avaraging over an ensemble of $C$
constructed SAW configurations is to be performed according to  
\begin{equation}
 \langle {\overline {O}} \rangle=\frac{{\sum_{j=1}^C W_{N_{{\rm SAW}}}^{{j}} {\overline {O}} ^{j} }} {Z_{N_{{\rm SAW}}}} \label{aver2}
 \end{equation}
 with  $W_{N_{{\rm SAW}}}^j$ being the weight of an $N_{{\rm SAW}}$-monomer chain in $j$th configuration of SAW as given by (\ref{weight1}),  $Z_{N_{{\rm SAW}}}=\sum_{j}^C W_{N_{{\rm SAW}}}^{{i}}$ and  ${\overline {O}}^{j}$ is the value obtained after averaging over all constructed DLA realizations on $j$th SAW configuration.

 {We applied averaging over $M=10^3$ DLA realizations and $C$ up to $10^4$ SAW configurations in our analysis below.}
}

\section{Results}\label{III}

First of all, we revisit the results for the DLA on a single particle seed constructed on a $D=2$ square lattice with the variable coagulation probability $p$ \cite{Ranguelov2012}.  

\begin{figure}[t!]
	\begin{center}
 \includegraphics[width=100mm]{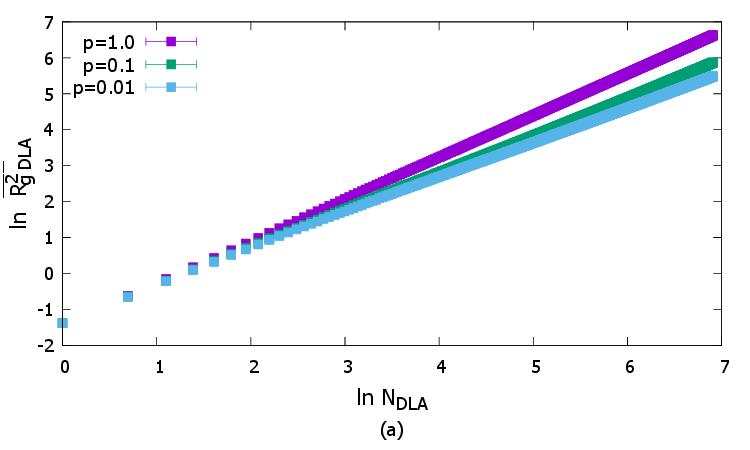}   
 \includegraphics[width=100mm]{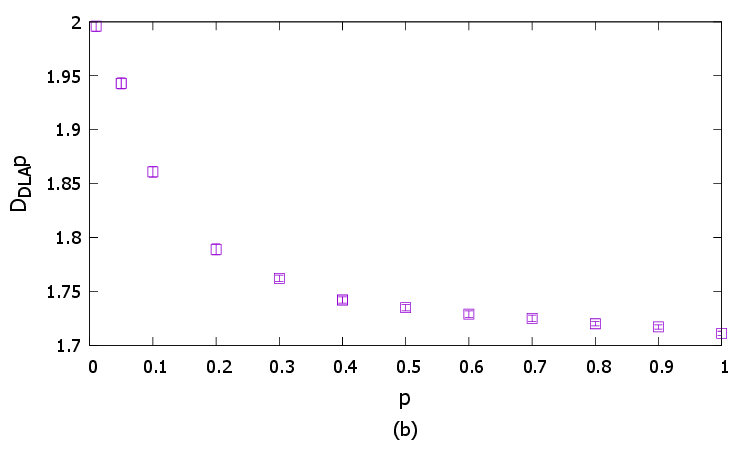}
		\caption {a) Average gyration radius of the DLA clusters obtained from a single particle seed, as the function of DLA size, $N_{{\rm  DLA}}$, at several values of $p$. b) Estimates for the fractal dimension of the DLA clusters as the function of $p$. }
		\label{RgDLAp}	
	\end{center}
\end{figure}
   
The gyration radius of DLA cluster is defined as 
\begin{equation}
R_{g\,DLA}^2=\frac{1}{N_{{\rm  DLA}}^2}\sum_{i=1}^{N_{{\rm  DLA}}} \sum_{j=1}^{DLA} ( (X_i-X_j)^2+(Y_i-Y_j)^2). \label{Rdef}
\end{equation}
It is subsequently averaged over the ensemble of DLA realizations, according to Eq. (\ref{aver1}), providing an average value ${\overline {R^2_{g}}}_{{\rm  DLA}}$. The data obtained  are shown in Fig. \ref{RgDLAp}a. Reduction of the $p$ value below $1$ gives more opportunity for particles to diffuse towards the center of a cluster. This results in formation of a more densely packed structures as compared with a standard DLA, $p=1$, see Fig.~\ref{cluster}. It is clearly seen from Fig. \ref{RgDLAp}a, that the value of  ${\overline {R^2_{g}}}_{{\rm  DLA}}$ decreases with the decrease of $p$. 
{ To evaluate the values of fractal dimensions $D_{{\rm  DLA}}$ of generated clusters, the linear least-square fits are performed. To this end we estimate the lower cutoff for the number of particles $N_{{\rm  DLA}}^{min}$ at which the correction to scaling terms become irrelevant. The linear fits for the average gyration radius is used 
\begin{equation}
\ln  {\overline {R^2_{g}}}_{{\rm  DLA}}  = A+ 2/D_{{\rm  DLA}} \ln N_{{\rm  DLA}}.
\end{equation}
The $\chi^2$ value (sum of squares of normalized deviation from the regression line) divided by the number of degrees of freedom, $DF$ serves as an estimate for the fit accuracy. An example is given in Table \ref{DLA05}}. The estimates for $D_{{\rm  DLA}}$ obtained as functions of $p$ are provided in Fig.~\ref{RgDLAp}b.  
\begin{table}[t!]
\begin{center}
    \begin{tabular}{|c|c|c|c|}
    \hline
        $N_{{\rm  DLA}}^{min}$ & $D_{{\rm  DLA}}$ & $A$ &  $\chi^2/DF $ \\ \hline
       5 & 1.743 (2) & -1.520(3) &   1.7296 \\ \hline	
       10 &  1.742 (2)  & -1.523(3) &  1.2942 \\ \hline	
       15  &  1.740 (1) & -1.526(2) & 0.9429 \\ \hline                  20 & 1.741 (1)  &    -1.525(2)   &  0.7421 \\ \hline
            \end{tabular}
    \caption{Results of the least-square fits for evaluation the fractal dimension $D_{{\rm  DLA}}$ when varying the lower cutoff $N_{{\rm  DLA}}^{min}$ for the case $p=0.5$  }
    \label{DLA05}
    \end{center}
\end{table}

\begin{figure}[b!]
 \begin{center}
\includegraphics[width=10.0cm]{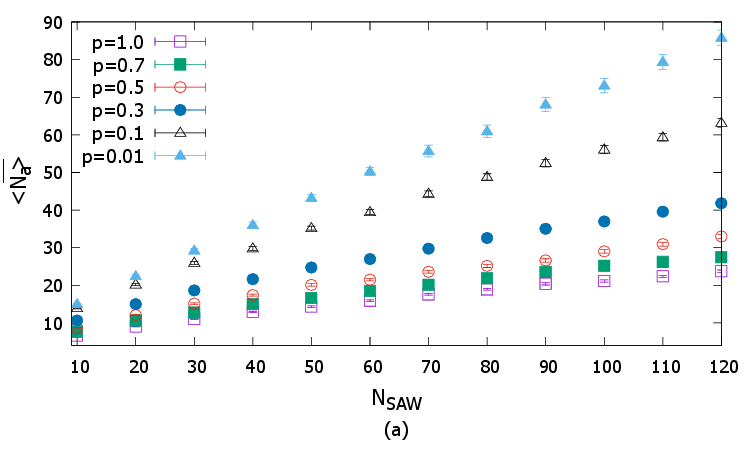}
\includegraphics[width=10.0cm]{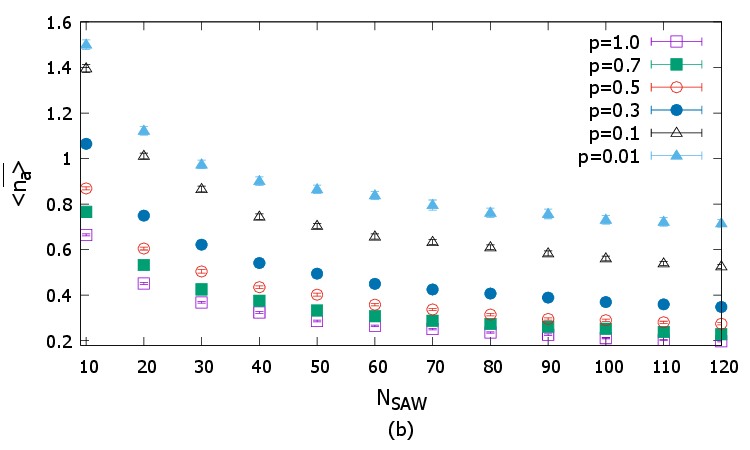}
\end{center}
\caption{\label{Na} Average number of particles $\langle \overline{ N_a} \rangle$ directly adsorbed on the SAW seed (a) and adsorption efficiency per single monomer $\langle \overline{n_a} \rangle$  (b), as  functions of $N_{{\rm SAW}}$ at several values of the coagulation probability $p$. }
\end{figure}

Now, we turn our attention to the formation of DLA on a seed in a form of SAW, see Fig. \ref{cluster}.
It is intuitively obvious that different perimeter sites of a such cluster are accessible to diffusive particles with uneven probabilities. The particles adsorbed on the polymer directly form a solvation shell, which causes a screening effect for the newly arriving particles. As the result, when the number of particles $N_a$ in a solvation shell reaches certain saturation value, the new particles are unable to reach the SAW core directly, but can be adsorbed by the solvation shell particles only.

Let us evaluate the size of the solvation shell of the polymer seed of size  $N_{{\rm SAW}}$, i.e. to estimate the number of particles $N_a$ adsorbed directly by the polymer seed. 
After performing double averaging, over an ensembles of SAW and DLA configurations, we obtain the results for  $\langle \overline {N_a} \rangle$ presented in Fig.~\ref{Na}a. As expected, with the decrease of the probability $p$,  the number of particles  $\langle \overline{N_a} \rangle$ directly adsorbed by a polymer flocculant increases. This results in formation of a more compact agglomerate, as illustrated  in Fig. \ref{cluster}b. We also introduce the normalized value $n_a = N_a/N_{{\rm SAW}}$, which characterizes the efficiency of direct adsorption of diffusive particles per single monomer of the SAW seed. By analysing behaviour of $\langle  {\overline n_a} \rangle$ as  function of $N_{{\rm SAW}}$ (Fig.~\ref{Na}b), one notices that the direct adsorption efficiency is  higher for shorter chains, and saturates at large values of $N_{{\rm SAW}}$.

\begin{figure}[t!]
	\begin{center}
	  \includegraphics[width=70mm]{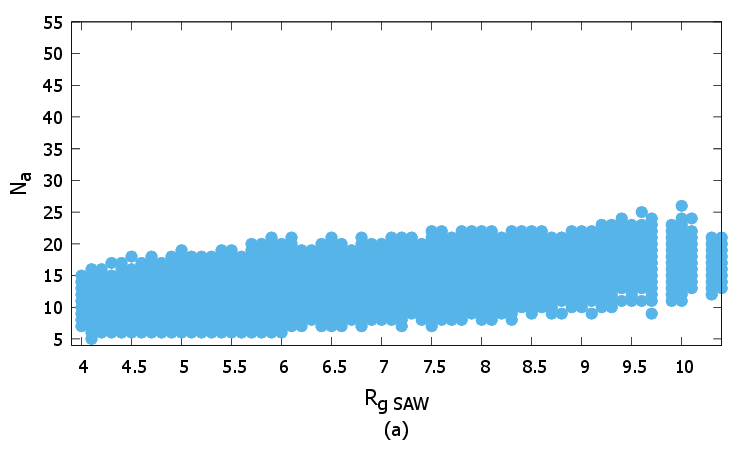}
        \includegraphics[width=70mm]{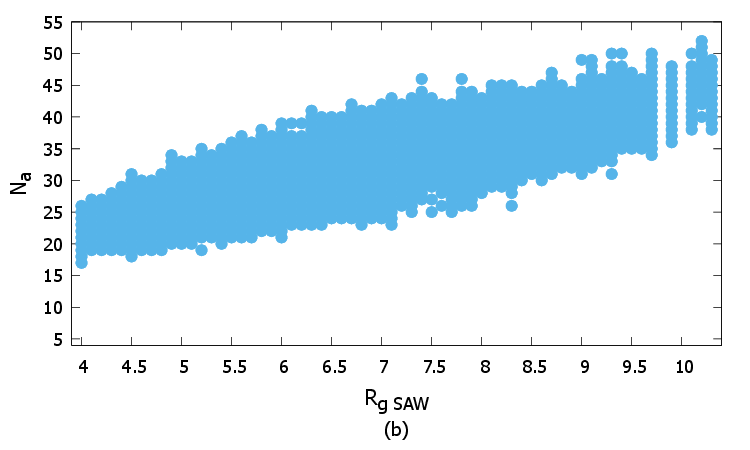}
  \caption {Correlation between the values of gyration radius $R_{g\, {\rm SAW}}$ of SAW seed with $N_{{\rm SAW}}=40$ monomers and the possible number of particles in solvation shell $N_a$ at $p=1.0$ (a) and $p=0.1$ (b).}
\label{Correlation}	
\end{center}
\end{figure}

\begin{figure}[t!]
	\begin{center}
	\includegraphics[width=70mm]{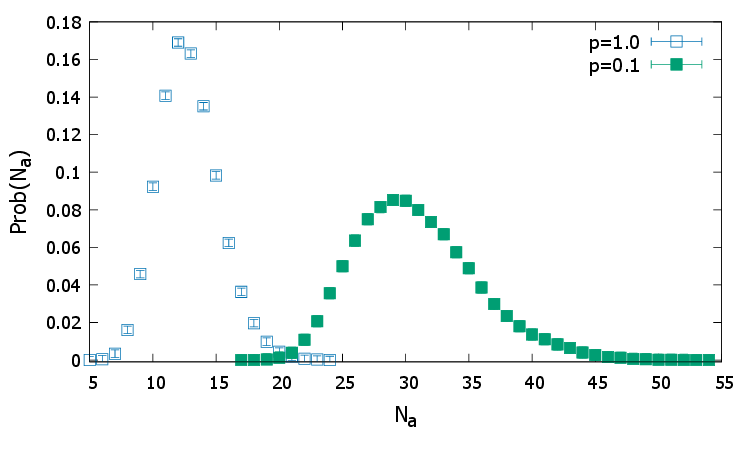}
  \caption { Probability distribution of a number of particles in solvation shell $N_a$ on a SAW seed with $ N_{{\rm SAW}}=40$ monomers with $p=1.0$ and $p=0.1$.}
\label{ProbNa}	
\end{center}
\end{figure}

{
Let us recall that the underlying SAW seed is itself a not-regular fractal structure with its effective size (gyration radius $R_{g\,{\rm SAW}}$) 
taking on a range of different values in an ensemble of possible SAW configurations. We analyzed the correlation between values of $R_{g\,{\rm SAW}}$ in given configuration of SAW and the numbers of particles in solvation shell $N_a$, observed in performing  different DLA realizations with this SAW configuration as a seed.  Corresponding results are presented on Fig. \ref{Correlation}. The smaller values of $R_{g\, {\rm SAW}}$ correspond to more compact SAW configurations with higher fraction of ``inner'' monomers, screened from incoming diffusive particles as compared with monomers positioned closer to the SAW outer perimeter. Indeed, the number of particles $N_a$ directly adsorbed to SAW perimeter is smaller for the case of compact SAW and increase gradually with an increase of  $R_{g\, {\rm SAW}}$. This effect is more pronounced for smaller $p$ values (Fig. \ref{Correlation}b), when direct adsorption of particles on SAW seed is dominant as compared with particle-particle coagulation, and thus is more ``sensitive'' to the subtle peculiarities of the underlying SAW.  Note that the probability distribution for $N_a$ as averaged over SAW and DLA realizations (see Fig. \ref{ProbNa}) is broader at smaller $p$  due to the same effect. 
}

{Another parameter of interest is the total number of bonds (contacts) $ N_{{\rm bond}}$  established between SAW and adsorbed diffusive particles. Since each the monomer of SAW can have contacts with more than one adsorbed particle, this quantity is not trivial and is expected to be larger than the  number of particles $N_a$ in a solvation shell. We prescribe labels  $v(n)$ to each $n$th monomer of the SAW ($n=1,\ldots,N_{{\rm SAW}}$) with the initial values of $v(n)=0$. When, during the process of diffusion, the $i$th  particle became adjacent to $n$th monomer and gets adsorbed, we increase the label $v(n)$ by one, so that:  
 \begin{itemize}
     \item 
 if  $X_i(t)=x_n\pm 1$ and $Y_i(t)=y_n$   then 
$ v(n)=v(n)+1$,
     \item 
 if $X_i(t)=x_n$  and $Y_i(t)=y_n\pm 1$  then 
$ v(n)=v(n)+1$
 \end{itemize}
  so that at the end $v(n)$ contains the total number of contacts established by $n$th monomer with adsorbed particles and $N_{{\rm bond}}=\sum_{n}v(n)$. }

\begin{figure}[b!]
 \begin{center}
\includegraphics[width=10.0cm]{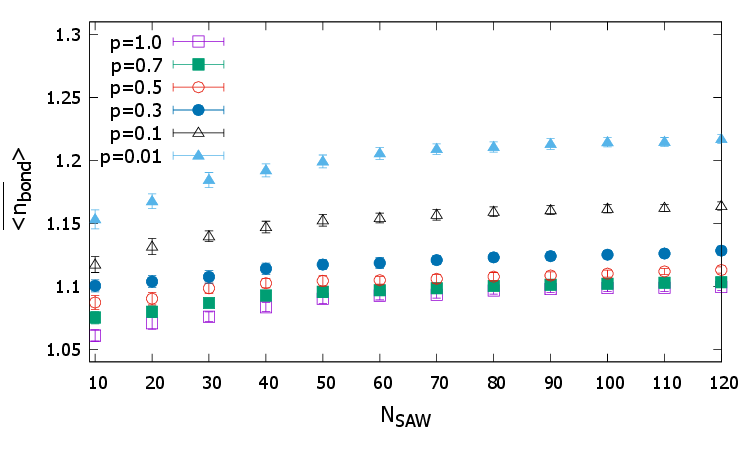}
\end{center}
\caption{\label{Nbond} The chelation abilities $\langle \overline{n_{bond}} \rangle$ of polymer chain as functions of $N_{{\rm SAW}}$ at several different values of the coagulation probability $p$.}
\end{figure}

{We introduce the value $n_{{\rm bond}}=N_{{\rm bond}}/N_{a}$, which  characterizes the chelation efficiency of a polymer chain and is also of interest from the point of view  of wastewater cleaning: the larger is the average number of contacts per single monomer, the stronger is the ability of a polymer flocculant to ``hold'' the particles, which are already adsorbed. At each fixed value of $p$,  $\langle {\overline {n_{{\rm bond}}}}\rangle$ is found to increase with the length of a polymer chain (see Fig. \ref{Nbond}) and gradually reaches its saturation value. The larger polymer macromolecules are thus found to be more effective in ``keeping'' the directly adsorbed impurity particles, and this efficiency increases with decreasing the parameter $p$.}

\begin{figure}[t!]
	\begin{center}		
 \includegraphics[width=100mm]{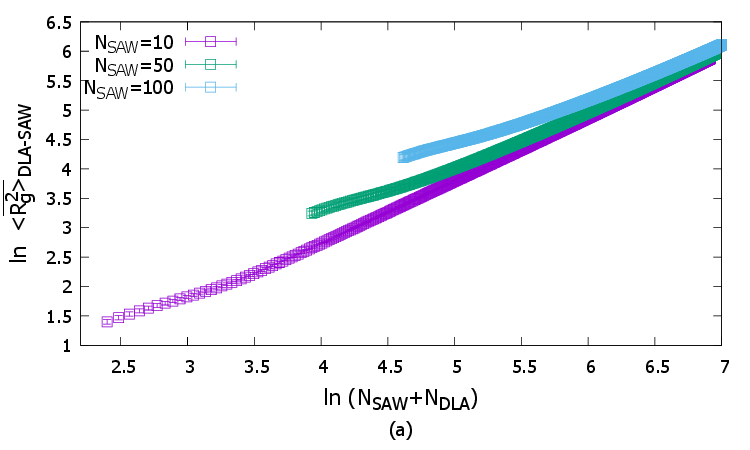}
   \includegraphics[width=100mm]{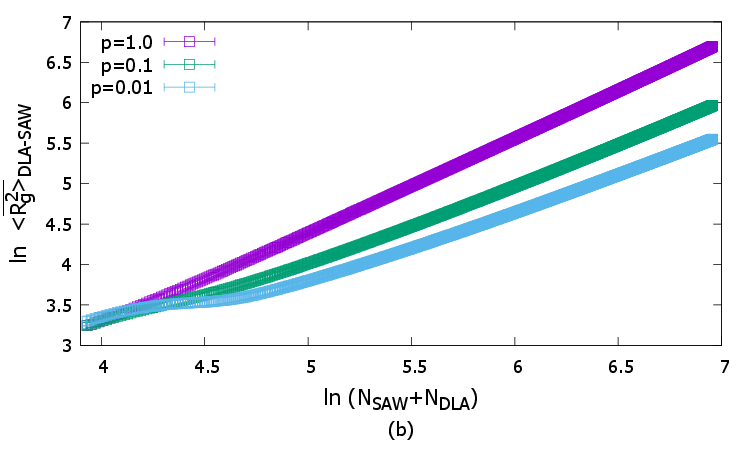}
		\caption {Average gyration radii of DLA-SAW aggregates,  a) At $p=0.1$ and several fixed values of parameter $N_{{\rm SAW}}$; b) At $N_{{\rm SAW}}=50$ and several fixed values of $p$.}
		\label{RgDLAeps}	
	\end{center}
\end{figure}

\begin{figure}[t!]
	\begin{center}		
 \includegraphics[width=100mm]{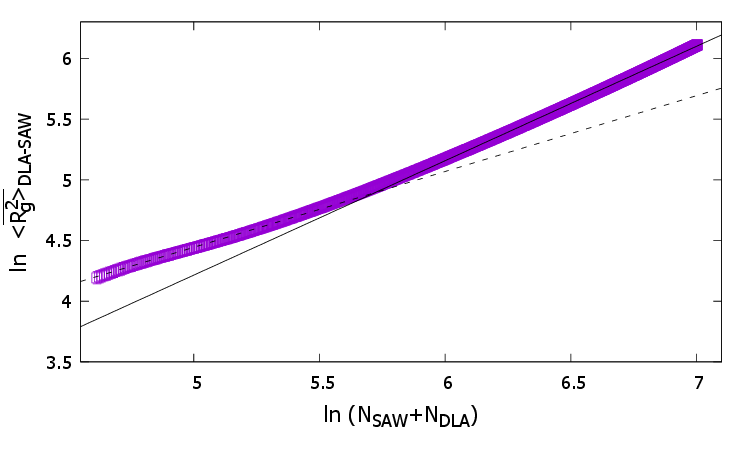}
		\caption {Average gyration radii of DLA-SAW aggregates at $p=0.1$ and  $N_{{\rm SAW}}=100$. Dashed line represents results of least-square fitting to the form (Eq. (1)) with $1/D=0.35(4)$ (solvation regime), solid line represents results of the least-square fitting to the form (Eq. (1)) with $D_{{\rm  {\rm  DLA-SAW}}}=1.868(5)$ (DLA regime). } 
		\label{RgDLAtest}	
	\end{center}
\end{figure}

Now we proceed to analyze the structure of agglomerates of $N_{{\rm SAW}}+N_{{\rm  DLA}}$ particles, as given by their fractal dimension. 
The scaling of the gyration radii $\langle {\overline{R^2_g}}\rangle_{{\rm  {\rm  DLA-SAW}}} $  in a range of sizes of the SAW seed $N_{{\rm SAW}}$ is analyzed next. This is related to a real life situations with various degree of pollution levels and of a pollutant coagulation ability. The data obtained are presented in both frames of Fig.~\ref{RgDLAeps}.
{ Here, we observe what we believe is the crossover between the solvation and DLA regimes, which is especially evident at small $p$, e.g. $p=0.1$. In this case, particles prefer to deposit theirselves on a SAW rather than to aggregate. Therefore, at low $N_{{\rm DLA}}<2N_{{\rm SAW}}$, the process is dominated by solvation (screening) of a SAW by particles, characterized by very slow increase of effective size $\langle {\overline{R^2_g}}\rangle_{{\rm  {\rm  DLA-SAW}}} $  with  $N_{{\rm  DLA}}$. By applying least-square fitting of data in this region to the form (Eq. (1)), we obtained  $1/D=0.35(3)$ (the dashed line in Fig. \ref{RgDLAtest}).  At $N_{{\rm DLA}}\sim 2N_{{\rm SAW}}$, a SAW is completely screened, and since then a normal DLA starts, where the screened SAW plays a role of its seed. At sufficiently large $N_{DLA}$, the DLA scaling with $D_{{\rm DLA-SAW}}(p=0.1)=1.868(5) $ is retrieved (the result of least-square fitting of data in this region to the form (Eq. (1)) are presented with solid line on Fig. \ref{RgDLAtest}), see also Table \ref{RN50p01}). The crossover is seen for all $N_{SAW}$ being examined, and it occurs at $N_{{\rm DLA}}\sim 2N_{{\rm SAW}}$ in all cases (as can be seen from curves of  Fig.~\ref{RgDLAeps}a). At yet smaller $p=0.01$, solvation regime has a non-linear dependence of $\langle {\overline{R^2_g}}\rangle_{{\rm  {\rm  DLA-SAW}}} $ on $N_{{\rm DLA}}$ due to strongly suppressed DLA in favour of screening the SAW, whereas at higher $p>0.1$ it gradually disappears, as in this case the screening of a SAW loses its priority over a DLA, see, respective curves in  Fig.~\ref{RgDLAeps}b.}

\begin{table}[b!]
\begin{center}
    \begin{tabular}{|c|c|c|c|}
    \hline
        $N^{min}_{{\rm  DLA}}$ & $D_{{\rm  {\rm  DLA-SAW}}}$ & $A$ &  $\chi^2/DF $ \\ \hline
       50 & 1.92(2) & -1.13(2) & 3.5344    \\ \hline		
       100  &  1.91(1)    &  -1.18(1)  &  2.3247    \\ \hline
       150 &  1.896(8) & -1.226(7) & 1.5064 \\ \hline
       200 & 1.896 (6) & -1.261(6) &  1.1872\\ \hline
       250 & 1.875 (6)  & -1.291(7) & 1.0922\\ \hline
       300 & 1.870 (5) & -1.303(6) & 0.9038\\ \hline
       350 & 1.868(5) & -1.308(5) &  0.7565\\ \hline
            \end{tabular}
    \caption{ Results of the least-square fitting for evaluation the fractal dimension $D_{{\rm  {\rm  DLA-SAW}}}$ with varying the lower cutoff $N_{{\rm  DLA}}^{min}$ at $N_{{\rm SAW}}=50$ and $p=0.1$. }
    \label{RN50p01}
    \end{center}
\end{table}




Let $M$ denote the number of realizations of DLA processes (which is 1000 in our case) on a fixed SAW configuration. For all the monomers $n=1,\ldots,N_{{\rm SAW}}$  we sum up the number of times $k(n)$, when  diffusive particle was absorbed to this monomer. In such a way, a weight $w_n=k(n)/M$ is prescribed to each monomer.
Such distributions of the hitting points are called the "harmonic measure" \cite{Kakutani1944},
which can be represented within the multifractal concepts. 
When studying physical processes on complex fractal objects, one often encounters the  situation of coexistence of a family of singularities, each associated with a set of different fractal dimensions \cite{Stanley1988}; combination of two fractal growth processes, as in our case, is expected to lead to multifractal features as well \cite{Blavatska2008a}.

\begin{figure}[t!]
	\begin{center}
	\includegraphics[width=100mm]{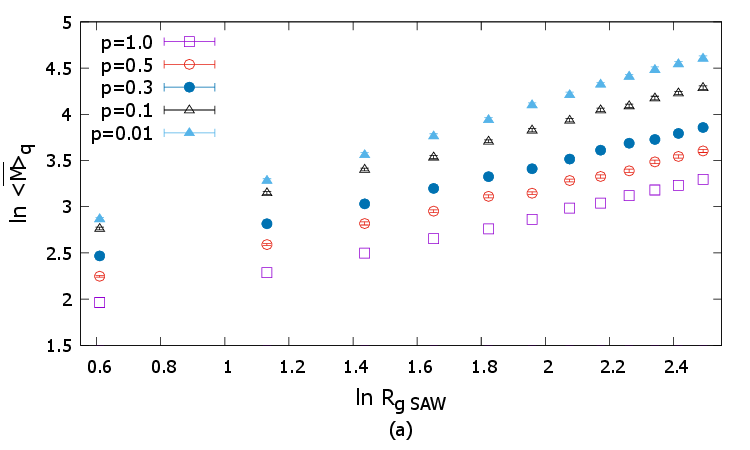}
        \includegraphics[width=100mm]{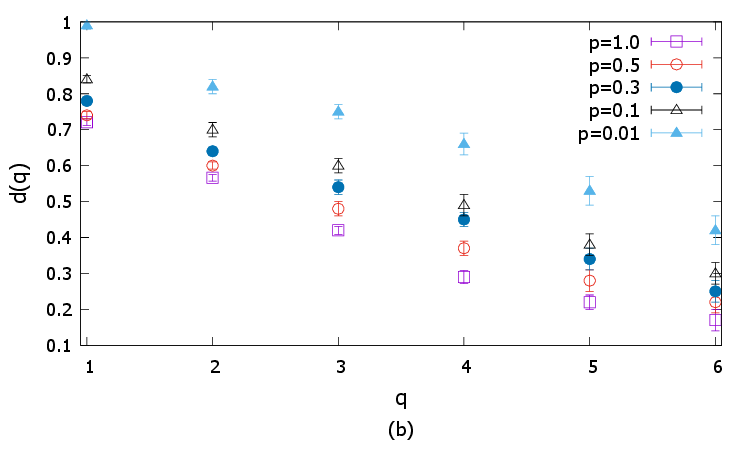}
		\caption {a) Average moments $\langle M(q)\rangle$ (Eq.(\ref{av_mom})) at $q=1$ and several values of $p$ as functions of $R_{g\,SAW}$ in double logarithmic scale. b) A spectrum of multifractal exponents $d(q)$  as function of $q$ at several values of $p$.}
		\label{times}	
	\end{center}
\end{figure}

The different moments of the distribution of observables scale independently, which is usually referred to as multifractality \cite{Mandelbrot1974}. The multifractal spectrum can be used to
provide information on the subtle geometrical properties
of a fractal object, which cannot be fully described by
its fractal dimensionality. Indeed,  the growth
probability distribution in DLA clusters is a typical example of multifractal phenomena \cite{Witten1981,Witten1983,Romo2013}.

In our case, the multifractal moments can be defined as
\begin{equation}
M(q)=\sum_{n=1}^{N_{{\rm SAW}}}w^q(n) \label{mom}
\end{equation}
When averaged over different configurations of the constructed SAW clusters and realizations of DLA, they scale with the gyration radius $R_g$ of an underlying SAW core according to:
\begin{equation}
\langle \overline{M(q)} \rangle=R_{g\,SAW}^{d(q)} \label{av_mom}
\end{equation}
with the exponents $d(q)$ being characterized by a non-linear dependence on $q$. To estimate the numerical values
of $d(q)$ on the basis of data obtained by us (see Fig. \ref{times}), the least square fitting was used.
At $q = 0$ we just count the number of sites of the cluster of linear size R, and thus $d(0)$ corresponds to the fractal dimension of the SAW trajectory for each $p$. At $q>0$, the set of $d(q) $ is found to be non-trivial and dependent on $p$. The obtained spectrum at several values of $p$ are given on Fig. \ref{times}b). At each $q$, the values of $d(q)$ increase with decreasing the parameter $p$. This demonstrates an increase of non-uniformity distribution of harmonic measure (the hitting probabilities of underlying SAW trajectory by incoming diffusive particles are distributed more with decreasing $p$). 

\section{Conclusions}

In this paper we mapped the process of coagulation-flocculation, which is important from the environmental point of view for cleaning wastewater, onto the physical model of the DLA of $N_{{\rm  DLA}}$ particles that takes place on a seed represented by a SAW containing $N_{{\rm SAW}}$ monomers.
{Within this approach, the DLA particles  represent the impurities in suspension, which  may sediment in form of flocs (aggregates) either spontaneously (as a result of particle-particle coagulation with probability $p$) or due to addition of a special agent (flocculant). The adsorbing linear polymer chain, represented by SAW here, serves as a flocculant by establishing multiple bonds with adsorbed DLA particles.}

Despite the fact that both paradigms are well studied over a number of years, their combination has got less attention, especially in relation to this particular application. Computer simulations are performed by means of the pruned-enriched Rosenbluth Monte Carlo algorithm on a $D=2$ square lattice.

Both DLA and SAW processes generate clusters of complex fractal structure. We found that their combination in the $D=2$ space, at the special case of {$N_{{\rm SAW}}=N_{{\rm  DLA}}$}, leads to formation of a {flocculated} agglomerate with the fractal dimension  $D_{{\rm  {\rm  DLA-SAW}}}=1.618(5)$, which exceeds the fractal dimension of a pure SAW, but is smaller than $D_{{\rm  DLA}}$. The result is relevant in the respect of the further removal of the agglomerate containing impurities by means of membrane filtration or gel chromatography. We also evaluated the properties related to the adsorption efficiency of SAW, such as: the average size of solvation shell $\langle N_a \rangle$, adsorbing efficiency per monomer $\langle n_a \rangle$, and a number of adsorption bonds between the SAW and the DLA particles $\langle N_{bond} \rangle$ depending on the $N_{{\rm SAW}}$ and the coagulation probability $p$. In particular we found,  that the direct adsorption efficiency as related to a single monomer of SAW, is  higher for shorter chain length, and reaches the asymptotic saturation value as   $N_{{\rm SAW}}$ increases.

The effective dimension of DLA-SAW agglomerates is given by their gyration radii $\langle R_{g\, {\rm  DLA-SAW}}^2 \rangle$, it is analyzed in a wide range of $N_{{\rm SAW}}$ and $p$. 
A crossover between two characteristic regimes have  been observed  at $N_{{\rm  DLA}} \gg N_{{\rm SAW}}$. At $N_{{\rm  DLA}} \leq N_{{\rm SAW}}$, the incoming diffusive particles are preferentially attached to monomers along the perimeter of SAW and form a solvation shell around the polymer seed; at $N_{{\rm  DLA}} \gg N_{{\rm SAW}}$,  the underlying SAW seed is already well ``screened'' from new incoming particles so that they mainly coagulate with particles from a solvation shell and the scaling regime of DLA is retrieved.

Introducing the probabilities for the perimeter sites of underlying SAW cluster to be encountered by incoming DLA particles, we found the estimates for multifractal sets of exponents, governing  the  moments (\ref{av_mom}) and giving more subtle characteristics of agglomeration clusters constructed by combination of two  growth processes.

The study contains a wide range of adjustable parameters that can be tuned towards particular chemical setup, namely: (i) molecular weight and branching type of a flocculant; (ii) relation between the impurity-flocculant adsorption and impurity-impurity coagulation; (iii) spatial arrangement of flocculants, e.g. polymer brush of variable density, etc. Such extensions of this study are planned for a future.

\section*{Acknowledgements}

The work was supported by Academy of Finland, reference number 334244.

 \section*{References}
\bibliography{DLA-SAW}
\end{document}